\documentclass[final,5p,times,twocolumn]{elsarticle}
\usepackage{lineno,hyperref}
\usepackage{hyperref}
\usepackage{epstopdf}

\journal{Journal of \LaTeX\ Templates}

\begin{document}

\begin{frontmatter}

\title{The Hybrid MPGD-based photon detectors of COMPASS RICH-1}


\author[infn-trieste,ICTP]{J.Agarwala}
\author[torino]{M.Alexeev}                                
\author[aveiro]{C.D.R.Azevedo}
\author[trieste]{F.Bradamante}                                
\author[trieste]{A.Bressan}                           
\author[freiburg]{M.B\"uchele}
\author[trieste]{C.Chatterjee}
\author[torino]{M.Chiosso}                                    
\author[infn-trieste,ICTP]{A.Cicuttin}
\author[trieste]{P.Ciliberti}
\author[infn-trieste,ICTP]{M.L.Crespo}
\author[infn-trieste]{S.Dalla Torre}                   
\author[infn-trieste]{S.Dasgupta}
\author[infn-torino]{O.Denisov}                               
\author[prague]{M.Finger}                                 
\author[prague]{M.Finger Jr.}                                         
\author[freiburg]{H.Fischer}  
\author[infn-trieste]{M.Gregori}                           
\author[infn-trieste]{G.Hamar}
\author[freiburg]{F.Herrmann}
\author[infn-trieste]{S.Levorato}
\author[trieste]{A.Martin}
\author[infn-trieste]{G.Menon}
\author[alessandria]{D.Panzieri}
\author[trieste]{G.Sbrizzai}                                       
\author[freiburg]{S.Schopferer}
\author[prague]{M.Slunecka}
\author[liberec]{M.Sulc}
\author[infn-trieste]{F.Tessarotto\corref{1}}
\author[aveiro]{J.F.C.A.Veloso}
\author[infn-trieste]{Y.Zhao}
\cortext[1]{
	corresponding author, email: fulvio.tessarotto@ts.infn.it}
%
%
\address[infn-trieste]{INFN, Sezione di Trieste, Trieste, Italy}
\address[torino]{INFN, Sezione di Torino and University of Torino, Torino, Italy}
\address[aveiro]{I3N - Physics Department, University of Aveiro, Aveiro, Portugal}
\address[trieste]{INFN, Sezione di Trieste and University of Trieste, Trieste, Italy}
\address[freiburg]{Universit\"at Freiburg, Physikalisches Institut, Freiburg, Germany}
\address[infn-torino]{INFN, Sezione di Torino, Torino, Italy}
\address[prague]{Charles University, Prague, Czech Republic and JINR, Dubna, Russia}
\address[alessandria]{INFN, Sezione di Torino and University of East Piemonte, Alessandria, Italy}
\address[liberec]{Technical University of Liberec, Liberec, Czech Republic}

\address[ICTP]{Also at Abdus Salam ICTP, 34151 Trieste, Italy}
%





\begin{abstract}
Novel gaseous detectors of single photons for RICH applications have been
developed and installed on COMPASS RICH-1 in 2016.
They have a hybrid architecture consisting of two staggered THGEM layers
(one equipped with a CsI photoconverting layer) and a bulk Micromegas; 
they cover a total area of 1.4 m$^2$ and operate stably and efficiently.
They provide a single photon angular resolution of $\sim$ 1.8 mrad
and about 10 detected photons per ring at saturation.
The main aspects of their construction and commissioning,
their characterization and performance figures are presented.
\end{abstract}


\end{frontmatter}


\section{Introduction}

The RICH-1~\cite{Albrecht} detector of the COMPASS Experiment~\cite{COMPASS} at CERN SPS
is a large gaseous Ring Imaging Cherenkov Counter providing hadron identification in the
range of momenta between 3 and 60 GeV/c, over a large angular acceptance ($\pm$200 mrad),
at high rates.

It consists of a 3~m long $C_{4}F_{10}$ radiator, a 21~$m^2$ large focusing VUV mirror
surface and Photon Detectors (PDs) covering a total active area of 5.5~$m^2$.
Three photodetection technologies are used in RICH-1:  Multi Wire Proportional Chambers
(MWPCs) with CsI photocathodes, Multi Anode Photo-Multipliers Tubes (MAPMTs) and
novel Micro Pattern Gaseous Detectors (MPGDs) based PDs.

COMPASS RICH-1 was designed and built between 1996 and 2001 and is in operation since 2002.
It was originally equipped with MWPCs hosting 16 CsI-coated photocathodes, each having
an active area of about 600$\times$600 mm$^2$; in 2006, to cope with the high particle flux
of the central region, 4 of the 16 CsI-coated photocathodes were replaced by detectors
consisting of arrays of MAPMTs coupled to individual fused silica lens telescopes.

In parallel, an extensive R\&D program~\cite{THGEM_rd}, aimed to develop MPGD-based large
area PDs, established a novel hybrid technology combining MicroMegas (MM) and Thick Gas
Electron Multipliers(THGEMs): this configuration provides good stability for
large area PDs operating in harsh conditions too.

In 2016 COMPASS RICH-1 was upgraded by replacing 4 MWPCs-based PDs with
detectors resulting from the newly developed MM+THGEM hybrid technology:
for the first time a running experiment uses MPGD-based detectors of single photons.

\section{The novel Hybrid MPGD-based Photon Detectors}

\begin{figure}[!htb]
	\centering
	\includegraphics[scale=1.15]{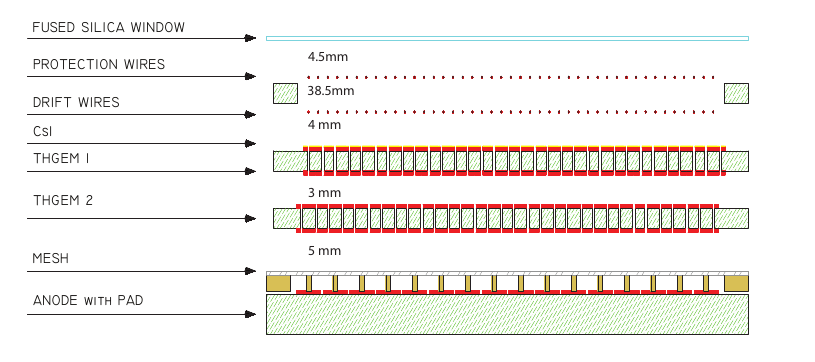}
	\caption{Sketch of the hybrid single photon detector: two THGEM layers are coupled to a MM.
		Drift and protection wire planes are shown. Image is not to scale.}
	\label{fig:hybrid}
\end{figure}

The novel Hybrid MPGD-based PD architecture, sketched in Fig.\ref{fig:hybrid},
consists in a combination of two layers of THGEM followed by a MM;
the top of the first THGEM is coated with a CsI film and acts as a reflective photocathode. 
In this configuration the feedback from photons generated in the multiplication
processes is suppressed by the presence of the THGEM layers and the large majority
of the ions from the MM multiplication are collected at the MM mesh.
The signal development time is $\sim$ 100 ns. 

Each of the four new COMPASS RICH-1 PDs has
600 $\times$ 600 mm$^2$ and is formed by two identical modules
(of 600 $\times$ 300 mm$^2$) arranged side by side.

\begin{figure}[!htb]
	\centering
	\includegraphics[scale=0.18]{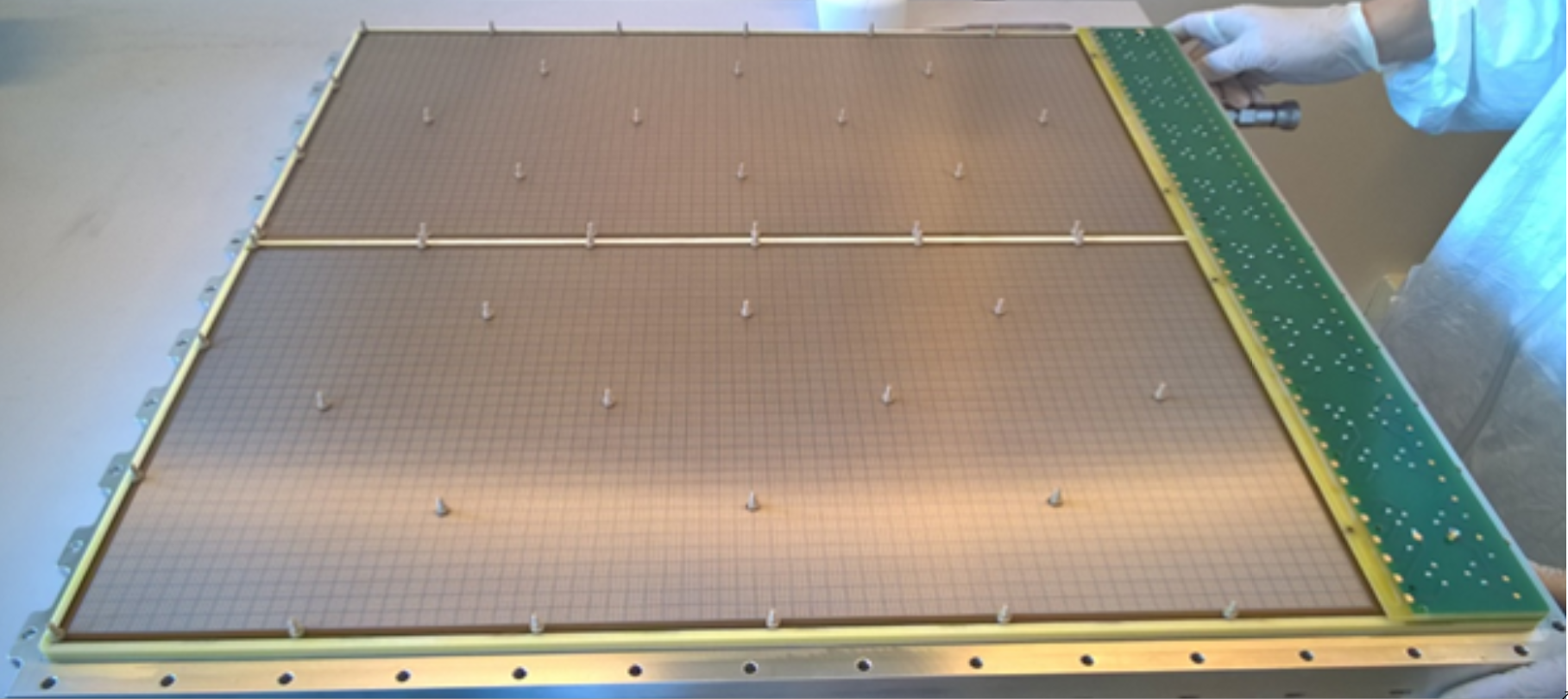}
	\caption{Two Micromegas mounted side by side in a PD.The pillars 
that preserve the distance between the micromesh and the THGEM above it are also visible.
	}
	\label{fig:MM}
\end{figure}

The MMs (Fig.\ref{fig:MM}) were produced at CERN using the bulk technology
\cite{Giomataris-2006-bulk-MM} 
on a custom, pad segmented anode; they have a 128 $\mu$m gap
and a square array of 300 $\mu$m diameter pillars with 2 mm pitch.

\begin{figure}[!htb]
	\centering
	\includegraphics[scale=0.08]{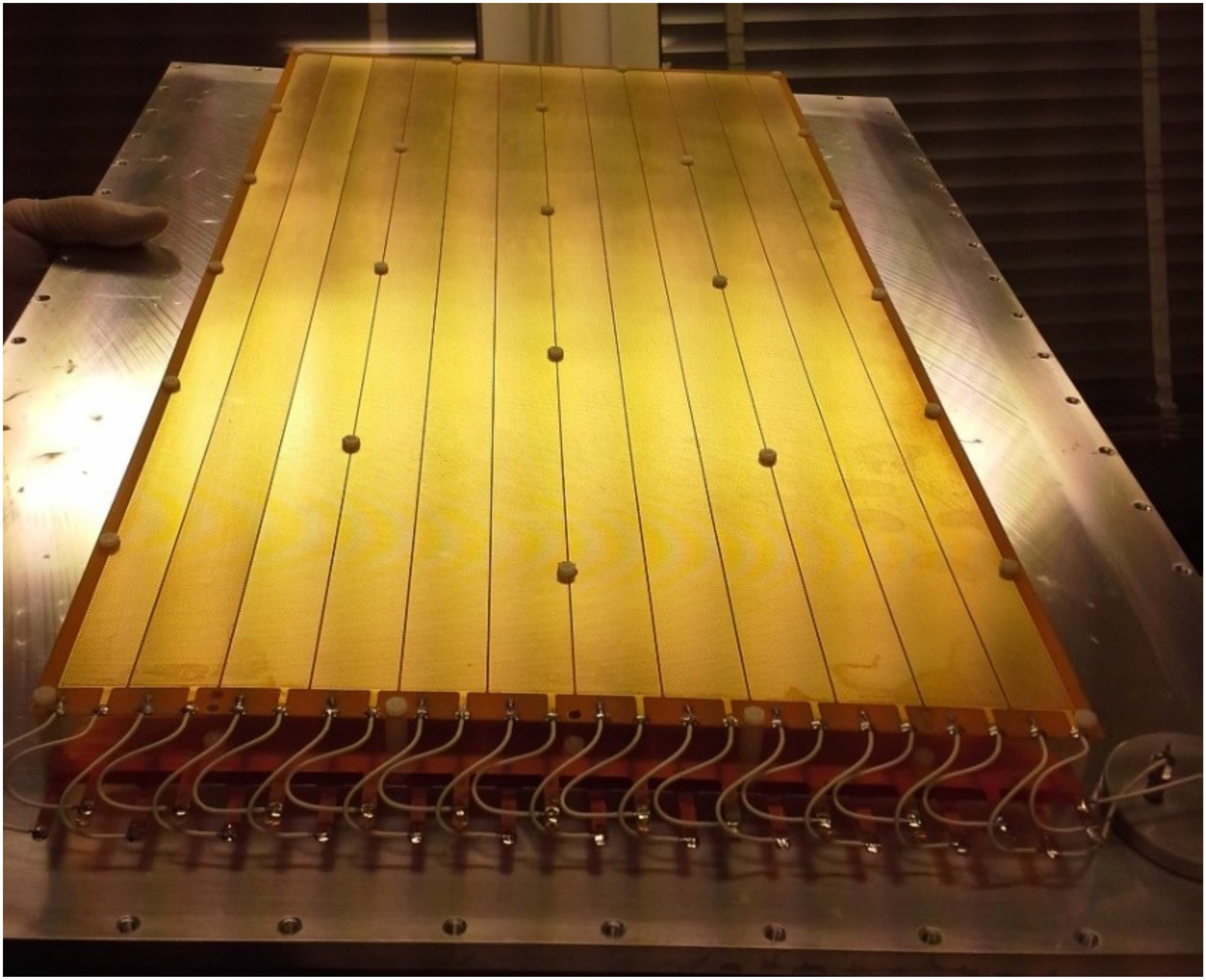}
	\caption{A Au coated THGEM ready for insertion in the CsI evaporation plant.}
	\label{fig:THGEM}
\end{figure}

The THGEMs (Fig.\ref{fig:THGEM}) are made from standard PCB material and their
geometrical parameters are: thickness = 470 $\mu$m
(400 $\mu$m dielectric, 2 $\times$ 35 $\mu$m copper), hole diameter = 400 $\mu$m,
pitch = 800 $\mu$m; the holes are rimless.
The THGEM top and bottom electrodes are segmented in 12 parallel
sectors, separated by 0.7 mm clearance, each biased via an individual
(1 G$\Omega$) resistor. 
The two THGEMs are staggered, providing maximal misalignment between
the two set of holes: this configuration increases the PD electrical
stability.

A specific protocol for production and validation of
the THGEMs has been applied, consisting in: preselection of raw PCB material
for homogeneous thickness, polishing after drilling with fine pumice powder,
cleaning with high pressure water and ultrasonic bath with a basic (PH11) solution,
detailed optical inspection, test of electrical strength, measurement of
gain uniformity and long test of discharge rates under illumination by X-rays.
The selected THGEMs were then coated with Ni (5$\mu$m) and Au (0.2$\mu$m)
(see Fig.\ref{fig:THGEM});
half of them were subject to a further coating with a 300 nm CsI layer
to become reflective photocathodes.
The quantum efficiency (QE) of the CsI photocathodes is measured inside the
evaporation plant after the coating process: the uniformity level is $\sim$ 3\%
r.m.s. within a photocathode and $\sim$ 10\% between different photocathodes.

\begin{figure}[!htb]
	\centering
	\includegraphics[scale=0.299]{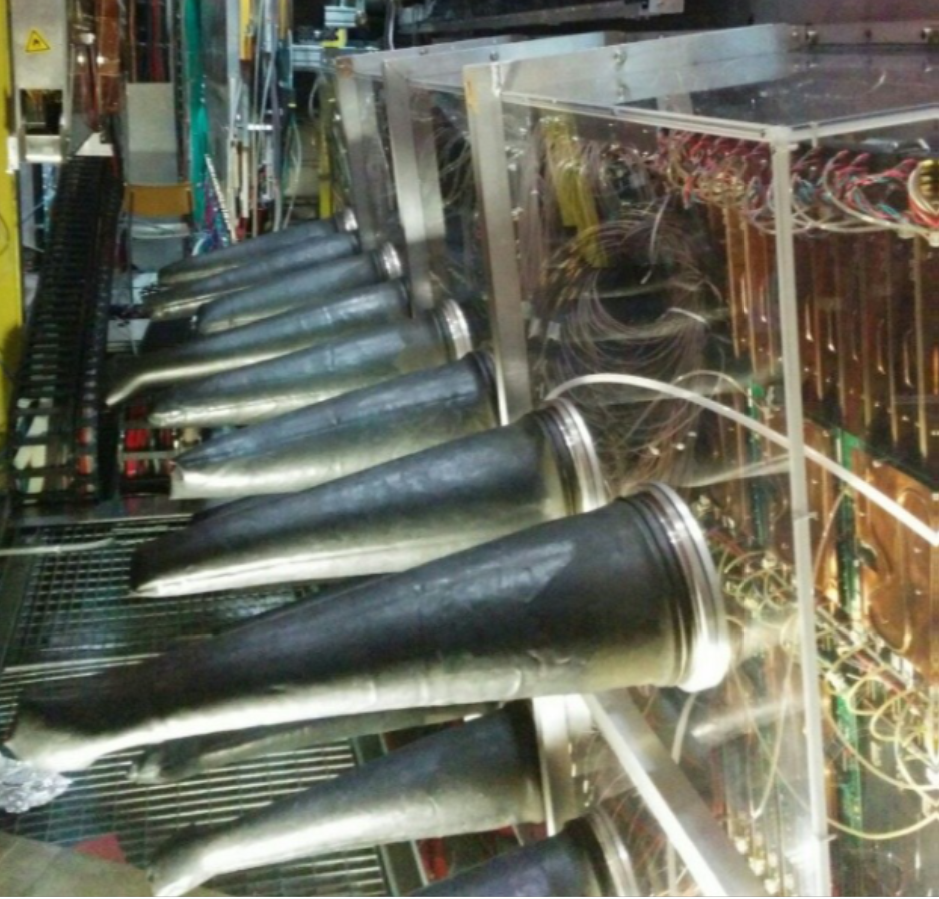}
	\caption{The glove box matching the RICH vessel mechanics used for the
		installation of the new PDs on the RICH.}
	\label{fig:glovebox}
\end{figure}

To preserve the QE all operations of transport and installation are performed
under controlled atmosphere, in dedicated glove-bexes (Fig.\ref{fig:glovebox}).

The hybrid PD anode is segmented in 7.5$\times$7.5 mm$^2$ pads with 0.5 mm interpad
clearence and each pad is biased at positive voltage ($\sim$ 620 V) via an individual
(470 M$\Omega$) resistor; the MM micromesh, being the only non-segmented electrode,
is kept at ground potential. 
This configuration prevents occasional
discharges from propagating to neighboring pads, limits the voltage
drop suffered by the pads surrounding a tripping one to about 2 V,
(corresponding to a gain drop $\sim$ 4\%) and allows restoring the
nominal voltage in few seconds. It also allows normal operation of the detector
even in the case one anodic pad is shorted to ground potential (a few cases
appeared during the runs 2016 and 2017, resulting in a total of less than 0.1\% dead MM area).
The signal is transmitted from the anode pad via capacitive coupling to a 
readout pad facing it, buried inside the anode PCB (at 70 $\mu$m distance from
the anode pad) and connected to the front-end board connector.
The resistive-capacitive pad scheme dumps the effects of discharges and protects
the front-end electronics.

\begin{figure}[!htb]
	\centering
	\includegraphics[scale=0.3]{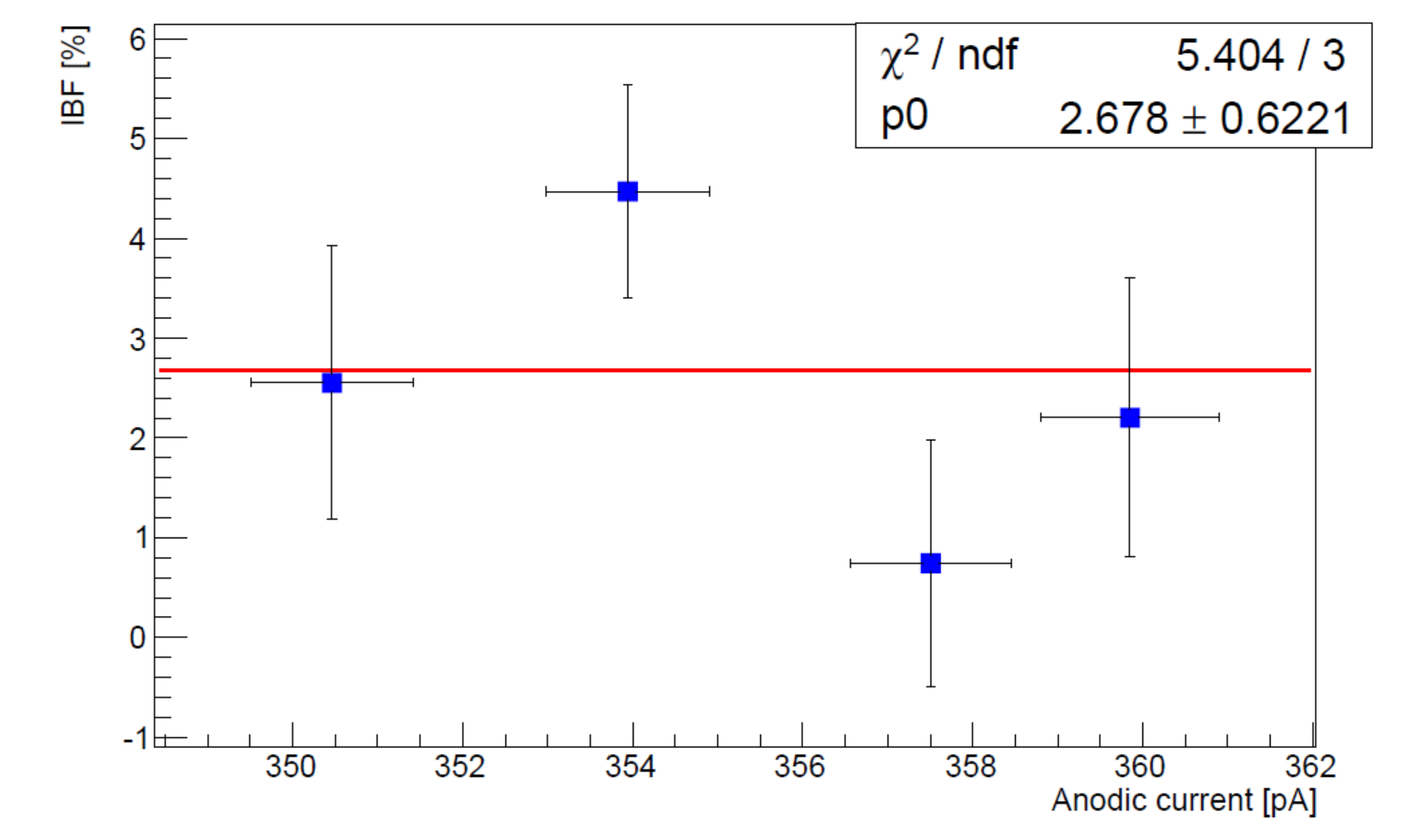}
	\caption{Dedicated measurement of ion back-flow fraction to the THGEM photocathode at four different
		values of the UV light intensities.}
	\label{fig:ibf}
\end{figure}

The novel hybrid PDs are operated on COMPASS RICH-1 with an Ar/CH$_4$ = 50/50 gas mixture.
The ion back-flow to the THGEM photocathode, in the standard operating conditions
has been measured to be $\le$ 3\% (see Fig.\ref{fig:ibf}).

\section{The commissioning of the MPGD-based PDs}

The new PDs have been installed on COMPASS RICH-1 during Spring 2016, commissioned
during the 2016 run, and operated stably and efficiently during the 2017 run too.
The high voltage is provided by commercial power suppliers (CAEN A1561HDN and A7030DP
HV modules, hosted in two SY4527 mainframes); each chamber is divided into 4 independent
sectors and has 9 different electrode types, each one having specific requirements.

The HV control system (custom made, using C++ and wxWidgets) monitors and records
at 1 Hz fequency the voltage and current values of all the 136 channels;
it counts the discharges (events with $\ge$
20 nA current increase) and readjusts automatically the specific voltage bias in case
the discharge rate exceeds the allowed limit. It also measures the variation of
environmental parameters (pressure and temperature) and provides automatic voltage
adjustment to compensate for it, in order to preserve the stability of the PD gain.
Discharges typically affect single
sectors only and the operating conditions are recovered in $\sim$ 10 s;
their rate is $\sim$ 1/h per chamber;
discharges in the two THGEM layers are fully correlated, while those observed in the MM
are mostly correlated with THGEM ones.
No high voltage power supply protection trips were observed during data taking.

The front-end electronics \cite{Neyret-2006-RICH-APV}, is based on the APV25-S1 chip,
and provides three amplitude samples per trigger for each channel.
Digitizer boards hosting 10-bit flash ADCs and FPGAs performing on-line zero suppression
with common-mode correction send the detector data to the COMPASS DAQ for data storage
and monitor. A cooling system using under-pressure water flow assures efficient removal
of the heat produced by the readout. The average noise level is $\sim$ 800
equivalent e$^-$ r.m.s, highly stable during the running periods. A clustering algorithm
is applied in the analysis to provide coordinates and amplitudes of "photon candidate"
clusters; the majority ($\ge$ 90\%) of clusters however receive contribution from
a single pad only.

\begin{figure}[!htb]
	\centering
	\includegraphics[scale=0.31]{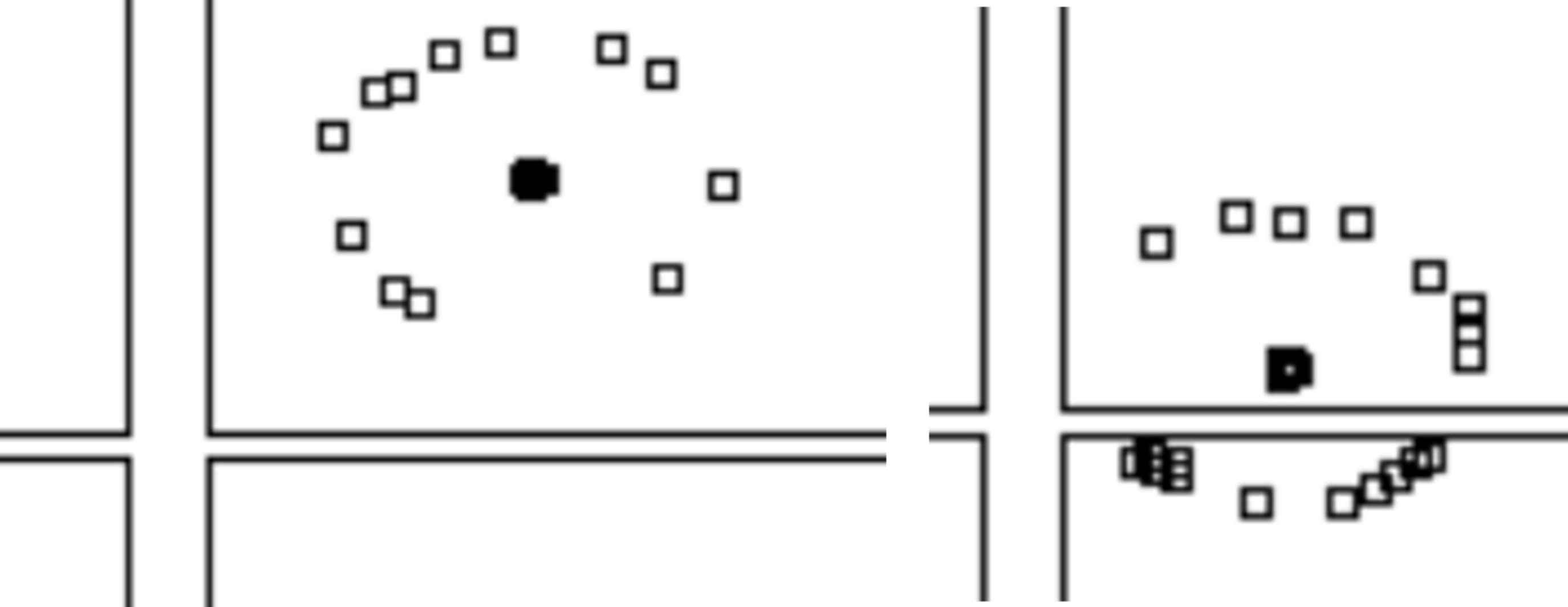}
	\caption{Examples of fully contained and shared rings}
	\label{fig:rings}
\end{figure}

The average effective gain of the 16 sectors was tuned to be the same (at 1\% level)
and to remain stable at a level of 5\% over several months of continuous operation.
The ring images provided by the novel detectors are clean and almost background-free:
typical examples are presented in Fig.\ref{fig:rings}, where a ring fully contained in
one of the new PDs (left) is shown together with a "shared" ring (right), namely a
ring with photons detected partly by the new PD and partly by the MAPMTs. In both
cases the reconstruction and identification efficiency is satisfactory.

\section{Preliminary performance results}

A preliminary characterization of the hybrid THGEM-MM PD response has been performed
from the analysis of data collected during two days of dedicated RICH calibration runs
in september 2017; during this period the radiator gas consisted in a mixture of
C$_4$F$_{10}$/N$_2$ $\sim$ 75/25.

\begin{figure}[!htb]
	\centering
	\includegraphics[scale=0.40]{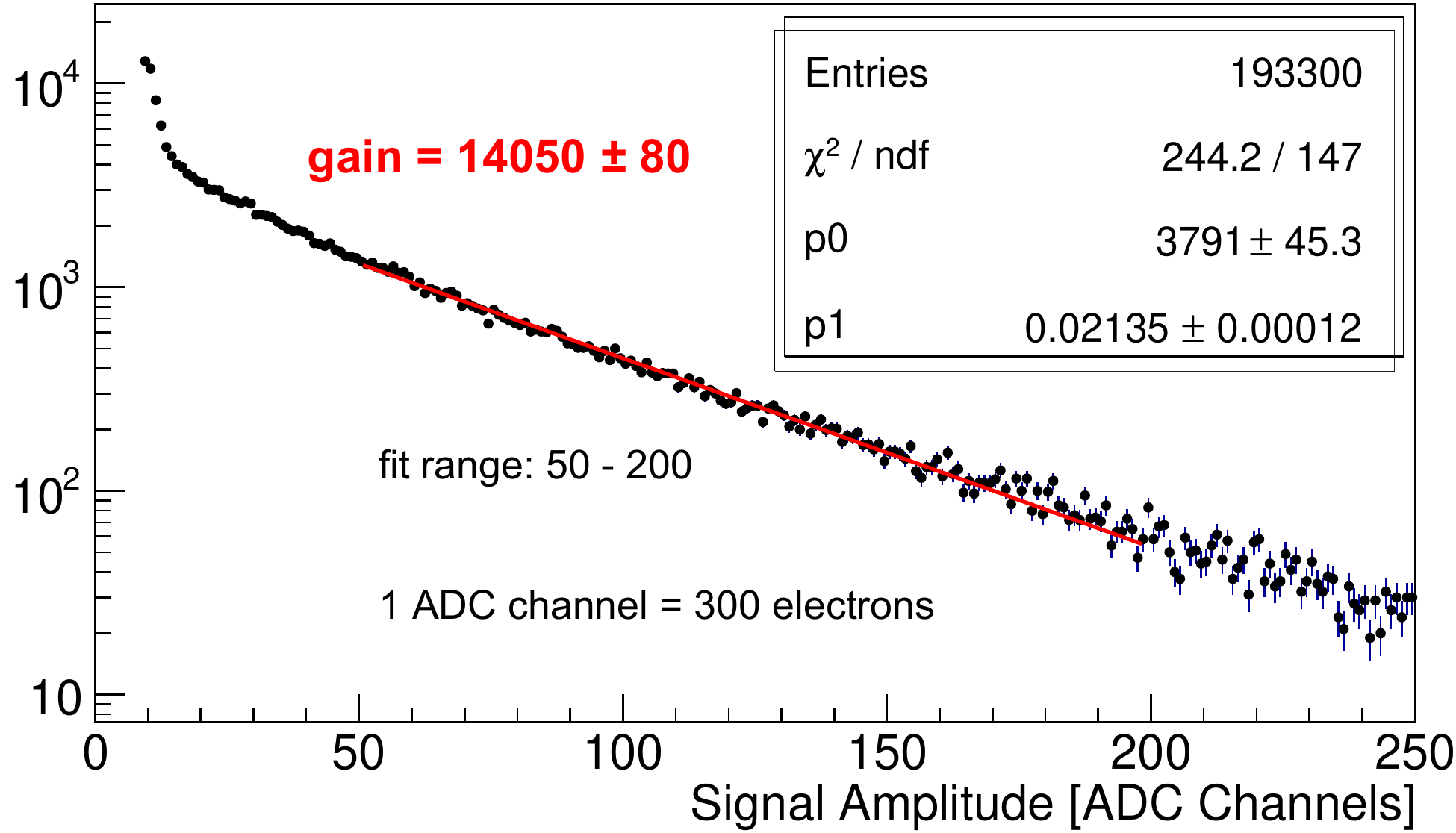}
	\caption{Signal amplitude distribution}
	\label{fig:gain}
\end{figure}

Photon candidate clusters contributing to the rings of identified particles are
selected to obtain pure samples of Cherenkov photoelectron signals: 
an example of their amplitude distribution is presented in Fig.\ref{fig:gain}
where the expected exponential behavior is
observed over more than two orders of magnitude. The extracted value for the
effective gain is $\sim$ 14000. The respective contributing factors from
the three layers (first THGEM, second THGEM and MM) of electron multipliers
to the effective gain are estimated to be $\sim$ 13, 9 and 120.

The single photoelectron detection efficiency, estimated from the measured
effective gain and the threshold applied to the signals, is $\ge$ 80\%; 
the level of background hits in the ring coronas,
originated by the electronic noise, is expected to be $\le$ 20\%:
the observed deviation of the hit amplitude distribution from a pure exponential
at very small amplitudes confirms this expectation.

\begin{figure}[!htb]
	\centering
	\includegraphics[scale=0.40]{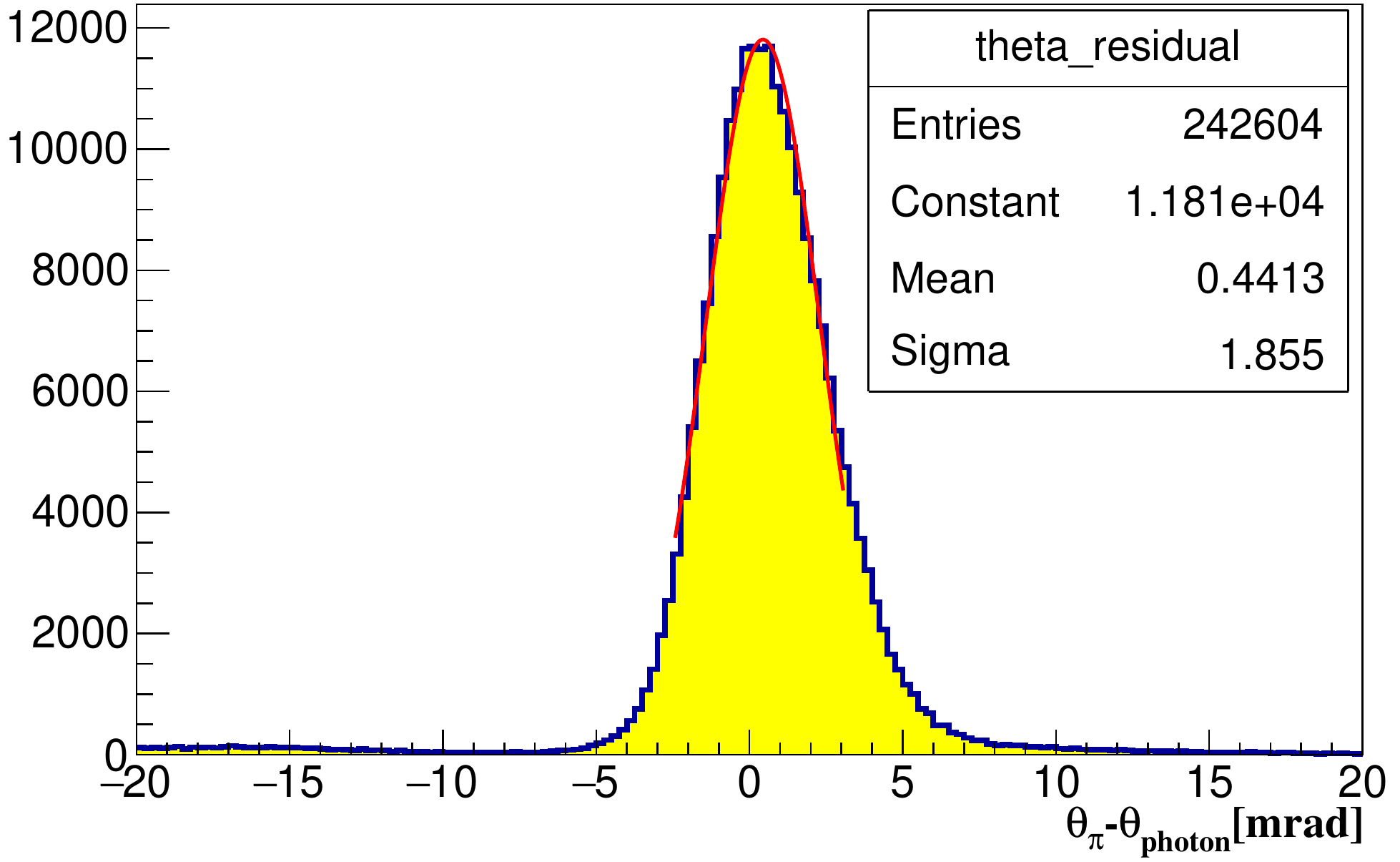}
	\caption{Detector angular resolution for single Cherenkov photons}
	\label{fig:angle}
\end{figure}

\begin{figure}[!htb]
	\centering
	\includegraphics[scale=0.2]{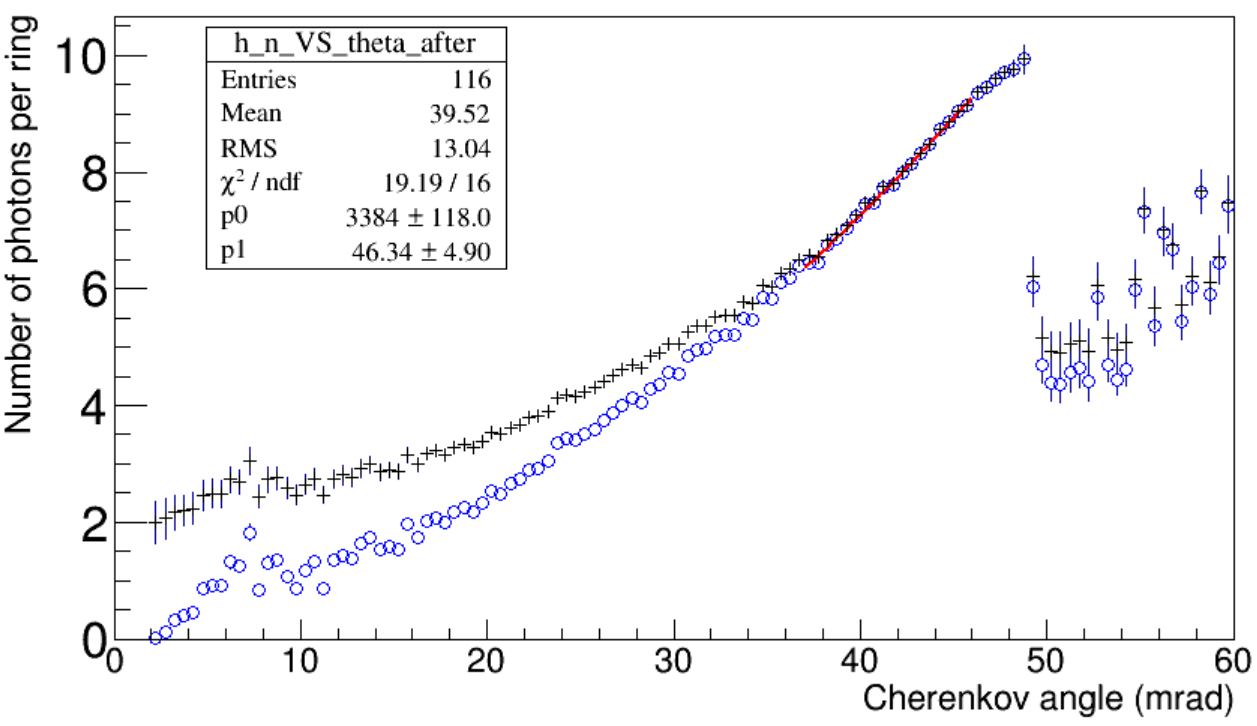}
	\caption{The number of detected photons per ring vs Cherenkov angle in a hybrid PD}
	\label{fig:N_ph}
\end{figure}

Selecting identified pion rings, the detector angular resolution for single
Cherenkov photoelectrons has been measured to be $\sim$ 1.85 mrad, as can be seen in 
Fig. \ref{fig:angle}, where the difference between the Cherenkov angle
calculated from the reconstructed momentum of the particle track and the measured
Chenekov angle for each photon candidate cluster is shown: this value fully matches
the expectation. 
The average number of detected photons per ring depends quadratically on the
Cherenkov angle, according to the Frank-Tamm equation: the observed number of
detected clusters shows the expected behavior, as can be seen in Fig. \ref{fig:N_ph},
where the points marked by crosses represent the measured quantity, while the
open circles represent the numbers corrected for the effect of the non
negligible probability of a statistical outcome of zero photons when the
average is very small. To increase the statistical accuracy of the estimate,
the shared rings are used too, provided at least half of the ring corona is
contained in the active area of the novel hybrid PD. 

A fit of the corrected distribution with a quadratic
(Frank-Tamm) + linear (random background proportional to the corona area)
function is then performed in the range of Cherenkov angles where high statistical
accuracy and small correction effects are present: since the quality of the fit is
good and the level of background obtained from the fit agrees with the expectation
a preliminary estimate for the number of detected photons for tracks at
saturation ($\beta$=1) can be reliably extracted.
The curve shown in Fig. \ref{fig:N_ph} provide a value of $\sim$ 13 hits at
a Cherenkov angle value of 55.2 mrad, which is the traditional reference
Cherenkov angle value at saturation for CsI photoconverter and a C$_4$F$_{10}$
radiator at s.t.p.;
the number of detected photons from the fit is $\sim$ 10.5 and the background
contribution is $\sim$ 2.5.

A complete characterization of the new detectors is still ongoing, but from the
preliminary results a clear indication of a very stable and reasonably high
effective gain, low noise level, good angular resolution and large photoelectron
detection efficiency is obtained.

\section{Conclusions}

Novel large area gaseous detectors of single photons, based on a hybrid combination
of THGEMs and Micromegas, have been developed and installed on COMPASS RICH-1 in 2016.
They operate stably and efficiently with an effective gain of $\sim$ 15000, a noise
level of $\sim$ 800 equivalent e$^-$ r.m.s., providing a single photon angular resolution
of $\sim$ 1.85 mrad and about 10 detected photons per ring at saturation. 

They represent a remarkable technological achievement, since gaseous PDs are the most
effective approach to instrument large surfaces with detectors of single photons
at affordable costs, and they have a very low magnetic sensitivity.

MPGD-based photon detectors are a promising option for future RICH applications too.


\section*{Acknowledgements}

The authors are grateful to the colleagues of the COMPASS
Collaboration for continuous support and encouragement.

This work is partially supported by the H2020 project
AIDA-2020, GA no. 654168.

\section*{References}


\end{document}